\documentclass[]{spie}  
\usepackage[]{subfigure}
\usepackage{dcolumn}
\usepackage{array}
\usepackage[pdftex]{graphicx}
\usepackage{wrapfig}
\usepackage{xcolor}
\usepackage{textcmds}
\DeclareGraphicsExtensions{.jpg,.pdf,.png}
\usepackage{ifthen}
\newboolean{useDraft}
\setboolean{useDraft}{true}
\usepackage{jay_extra}

\usepackage{epstopdf}
\usepackage{amsmath}
\usepackage{lscape} 
\usepackage[T1]{fontenc}
\usepackage{ae,aecompl}
\usepackage{remreset}
\usepackage{url}
\usepackage{wasysym}

\newcommand{\lcdm}{\mbox{$\Lambda$CDM}}

\newcommand{\des}{{\sc des}}

\def\sqdeg{\ensuremath{\mathrm{deg}^2}}

\def\uk{\ensuremath{\mu \mathrm{K}}}
\def\ukarcmin{\uk-arcmin}

\def\summnu{\ensuremath{\Sigma m_\nu}}
\def\neff{\ensuremath{N_\mathrm{eff}}}
\def\ombh{\ensuremath{\Omega_b h^2}}
\def\omch{\ensuremath{\Omega_c h^2}}
\def\spitzer{\textit{Spitzer}}
\def\herschel{\textit{Herschel}}
\def\xmm{\textit{XMM-Newton}}
\def\wmap{\textit{WMAP}}
\def\planck{\textit{Planck}}

\title{SPTpol: an instrument for CMB polarization measurements with the South Pole Telescope}

\author{
J.E.~Austermann\supit{2,3},  
K.A.~Aird\supit{20},  
J.A.~Beall\supit{4},  
D.~Becker\supit{4},  
A.~Bender\supit{6},  
B.A.~Benson\supit{1,21},  
L.E.~Bleem\supit{1,19},  
J.~Britton\supit{4},  
J.E.~Carlstrom\supit{1,5,19,21,22},  
C.L.~Chang\supit{1,5,21},  
H.C.~Chiang\supit{1,21},  
H-M.~Cho\supit{4},  
T.M.~Crawford\supit{1,22},  
A.T.~Crites\supit{1,22},  
A.~Datesman\supit{7},  
T.~de Haan\supit{6},  
M.A.~Dobbs\supit{6},   
E.M.~George\supit{8},  
N.W.~Halverson\supit{2,3,18},  
N.~Harrington\supit{8},  
J.W.~Henning\supit{2,3},  
G.C.~Hilton\supit{4},  
G.P.~Holder\supit{6},  
W.L.~Holzapfel\supit{8},  
S.~Hoover\supit{1,19},  
N.~Huang\supit{1,21},  
J.~Hubmayr\supit{4},  
K.D.~Irwin\supit{4},  
R.~Keisler\supit{1,19,21},  
J.~Kennedy\supit{6},  
L.~Knox\supit{23},  
A.T.~Lee\supit{8},  
E.~Leitch\supit{1},  
D.~Li\supit{4},  
M.~Lueker\supit{10},  
D.P.~Marrone\supit{17},  
J.J.~McMahon\supit{11},  
J.~Mehl\supit{1,21},  
S.S.~Meyer\supit{1,19,21,22},  
T.E.~Montroy\supit{13},  
T.~Natoli\supit{1,19},  
J.P.~Nibarger\supit{4},  
M.D.~Niemack\supit{4},  
V.~Novosad\supit{7},  
S.~Padin\supit{1}, 
C.~Pryke\supit{12},  
C.L.~Reichardt\supit{8},  
J.E.~Ruhl\supit{13},  
B.R.~Saliwanchik\supit{13},  
J.T.~Sayre\supit{13},  
K.K.~Schaffer\supit{14},  
E.~Shirokoff\supit{10},  
A.A.~Stark\supit{24},  
K.~Story\supit{1,19},
K.~Vanderlinde\supit{6},  
J.D.~Vieira\supit{10},  
G.~Wang\supit{5},  
R.~Williamson\supit{1,21},  
V.~Yefremenko\supit{5,7},  
K.~W.~Yoon\supit{4},  
O.~Zahn\supit{8}
\skiplinehalf
\supit{1} Kavli Institute for Cosmological Physics, Department of Physics, Enrico Fermi Institute, The University of Chicago, Chicago, IL 60637, USA
\skiplinehalf
\supit{2} Department of Astrophysical and Planetary Sciences, University of Colorado, Boulder, Colorado,80309, USA
\skiplinehalf
\supit{3} Center for Astrophysics and Space Astronomy, University of Colorado, Boulder, Colorado,80309, USA
\skiplinehalf
\supit{4} NIST, Boulder, CO 80305, USA
\skiplinehalf
\supit{5} High Energy Physics Division, Argonne National Laboratory, Argonne, IL 60439, USA
\skiplinehalf
\supit{6} McGill University, Montreal, Quebec, Canada
\skiplinehalf
\supit{7} Materials Science Division, Argonne National Laboratory, Argonne, IL 60439, USA
\skiplinehalf
\supit{8} University of California, Berkeley, 151 LeConte Hall Berkeley, CA 94720, USA
\skiplinehalf
\supit{10} California Institute of Technology, Pasadena, CA 91125, USA
\skiplinehalf
\supit{11} University of Michigan, Ann Arbor, Michigan, USA
\skiplinehalf
\supit{12} University of Minnesota, Minneapolis, MN 55455, USA
\skiplinehalf
\supit{13} Case Western Reserve University, Cleveland, Ohio 44106, USA
\skiplinehalf
\supit{14} School of the Art Institute of Chicago, Chicago, Illinois, 60603, USA
\skiplinehalf
\supit{17} Steward Observatory, University of Arizona, 933 North Cherry Avenue, Tucson, AZ
85721, USA
\skiplinehalf
\supit{18} Department of Physics, University of Colorado, Boulder, CO 80309
\skiplinehalf
\supit{19} Department of Physics, University of Chicago, 5640 South Ellis Avenue, Chicago, IL, USA 60637
\skiplinehalf
\supit{20} University of Chicago, 5640 South Ellis Avenue, Chicago, IL, USA 60637
\skiplinehalf
\supit{21} Enrico Fermi Institute, University of Chicago, 5640 South Ellis Avenue, Chicago, IL, USA 60637
\skiplinehalf
\supit{22} Department of Astronomy and Astrophysics, University of Chicago, 5640 South Ellis Avenue, Chicago, IL, USA 60637
\skiplinehalf
\supit{23} Department of Physics, University of California, One Shields Avenue, Davis, CA, USA 95616
\skiplinehalf
\supit{24} Harvard-Smithsonian Center for Astrophysics, 60 Garden Street, Cambridge, MA, USA 02138
}

\authorinfo{J.E.A.: E-mail: jason.austermann@colorado.edu}
 
\begin{document} 
  \maketitle 

\begin{abstract}
SPTpol is a dual-frequency polarization-sensitive camera that was deployed on the 10-meter South Pole Telescope in January 2012.  SPTpol will 
measure the polarization anisotropy of the cosmic microwave background (CMB) on angular scales spanning an arcminute to several degrees.   The polarization 
sensitivity of SPTpol will enable a detection of the CMB ``B-mode" polarization from the detection of the gravitational lensing of the CMB by large scale structure, and a detection or improved upper limit on a primordial signal due to inflationary gravity waves.  The two measurements can be used to constrain the sum of the neutrino masses 
and the energy scale of inflation.  These science goals can be achieved through the polarization sensitivity of the SPTpol camera and careful control 
of systematics.  The SPTpol camera consists of 768 pixels, each containing two transition-edge sensor (TES) bolometers coupled to orthogonal polarizations, and a total 
of 1536 bolometers.  The pixels are sensitive to light in one of two frequency bands centered at 90 and 150 GHz, with 180 pixels at 90 GHz and 588 pixels at 150 GHz.  
The SPTpol design has several features designed to control polarization systematics, including: single-moded feedhorns with low cross-polarization, 
bolometer pairs well-matched to difference atmospheric signals, an improved ground shield design based on far-sidelobe measurements of the SPT, and a 
small beam to reduce temperature to polarization leakage.  We present an overview of the SPTpol instrument design, project status, and 
science projections. 

\end{abstract}




\section{INTRODUCTION}
\label{sec:int} 

The South Pole Telescope (SPT) is a 10-meter mm-wavelength telescope at the geographic South Pole.  In November 2011, the SPT completed a 
2500~\sqdeg\ survey at 95, 150, and 220 GHz, the SPT-SZ survey, that has already led to significant results and new discoveries in three main 
areas: using the Sunyaev-Zel'dovich (SZ) effect to discover massive high-redshift galaxy clusters and constrain cosmological 
parameters\cite{staniszewski09, vanderlinde10, williamson11, benson11, reichardt12}, measurements of fine-scale anisotropy 
of the cosmic microwave background (CMB) and the gravitational lensing of it from large scale structure\cite{lueker10, shirokoff11, keisler11, reichardt11, vanengelen12}, 
and the discovery of strongly lensed high-redshift star forming galaxies\cite{vieira10}.  

In January 2012, the SPT was equipped with a new polarization-sensitive camera, SPTpol.   SPTpol is a dual-frequency
polarization-sensitive camera sensitive to 90 and 150 GHz that will continue the original SPT science goals, and, in addition, measure the 
polarization anisotropy of the CMB on angular scales from an arcminute to several degrees.  CMB polarization measurements 
are sensitive to multiple cosmological parameters that are related to open questions in fundamental physics, including the sum of the neutrino masses, 
the number of relativistic particle species at recombination, and the energy scale of inflation.  The sensitivity of the SPTpol camera is expected 
to enable new constraints on each through its measurement of the power spectrum of the CMB polarization anisotropy.  

This paper is organized as follows.  In Section \ref{sec:sci}, we discuss the science goals for SPTpol.  In Section \ref{sec:tel}, we describe the telescope 
and optics.  In Section \ref{sec:rec}, we describe the design of the SPTpol camera, including the detectors and readout.  In Section \ref{sec:obs}, we discuss the 
SPTpol survey strategy.  Finally, in Section \ref{sec:pro}, we present the details of the SPTpol receiver performance, on-sky characterization, 
and the science projections for the SPTpol survey.  


\section{Science Goals}
\label{sec:sci}  

SPTpol will make sensitive measurements of the temperature and polarization anisotropy of the CMB on angular scales ranging from
a few degrees to an arcminute, or angular multipoles between $\ell \sim $50 -- 10,000.   The polarization signal can be decomposed into 
what are commonly referred to as ``E-mode'' (gradient-like)  and ``B-mode'' (curl-like) signals; an allusion to the 'E' and 'B' field 
patterns in electromagnetism.  These patterns are sourced and/or induced by distinct physical properties which can be separated through this decomposition.
These measurements contain a wealth of information about the initial conditions, content, and evolution of the Universe, and can 
be used to constrain a host of fundamental cosmological parameters, 
including: the mass and number of neutrinos, the primordial power spectrum, the energy scale of inflation, and the density and 
equation of state of dark energy.  In this section, we will discuss the science goals of the SPTpol experiment, and 
the projected cosmological constraints.  

\subsection{B-Mode Polarization Science}

\begin{figure}[t]
\centering
\begin{minipage}[c]{0.49\textwidth}
\includegraphics[trim=0pt 0pt 0pt 0pt,width=0.99\textwidth]{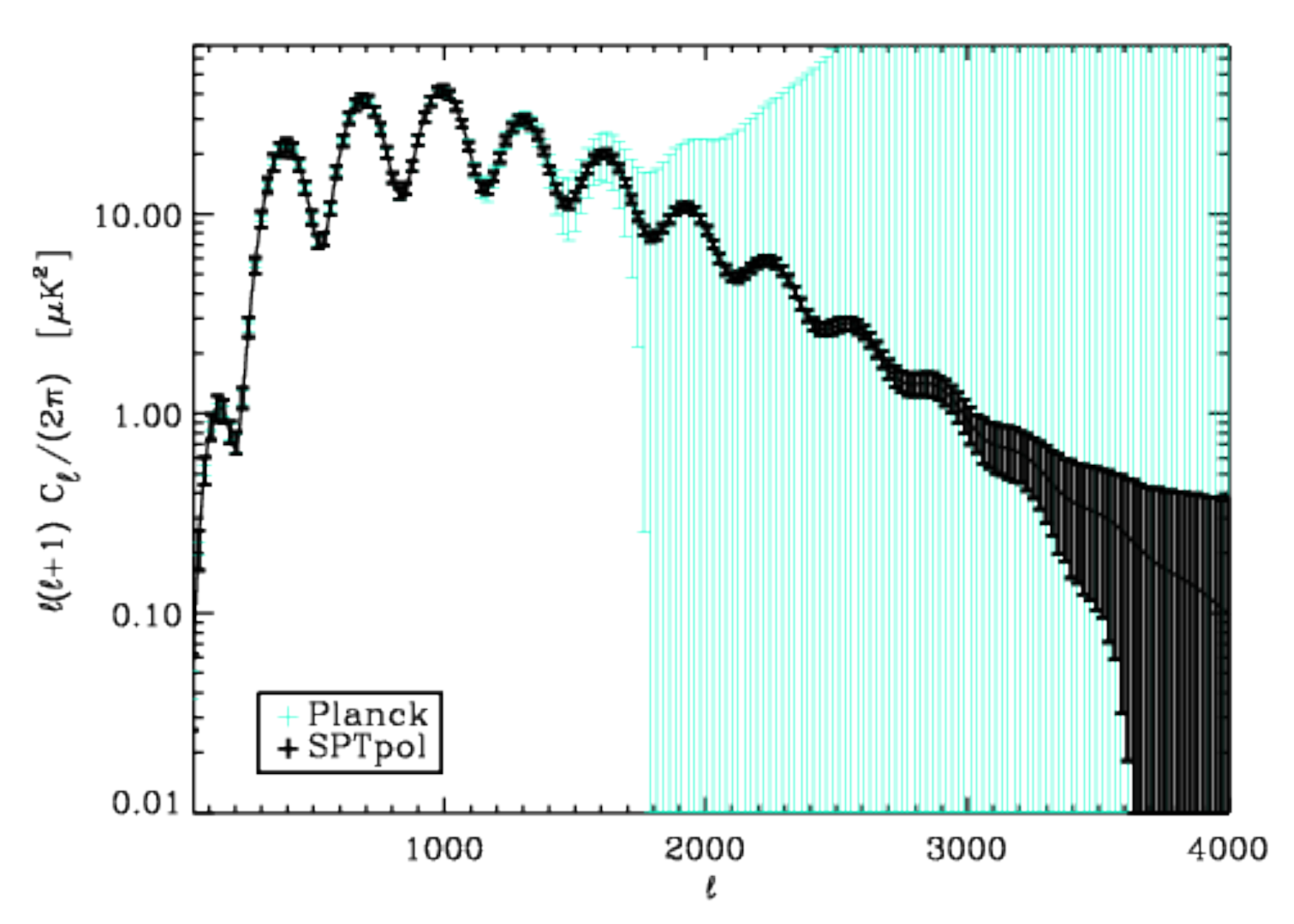}
\end{minipage}
\begin{minipage}[c]{0.49\textwidth}
\includegraphics[trim=0pt 0pt 0pt 0pt,width=0.99\textwidth]{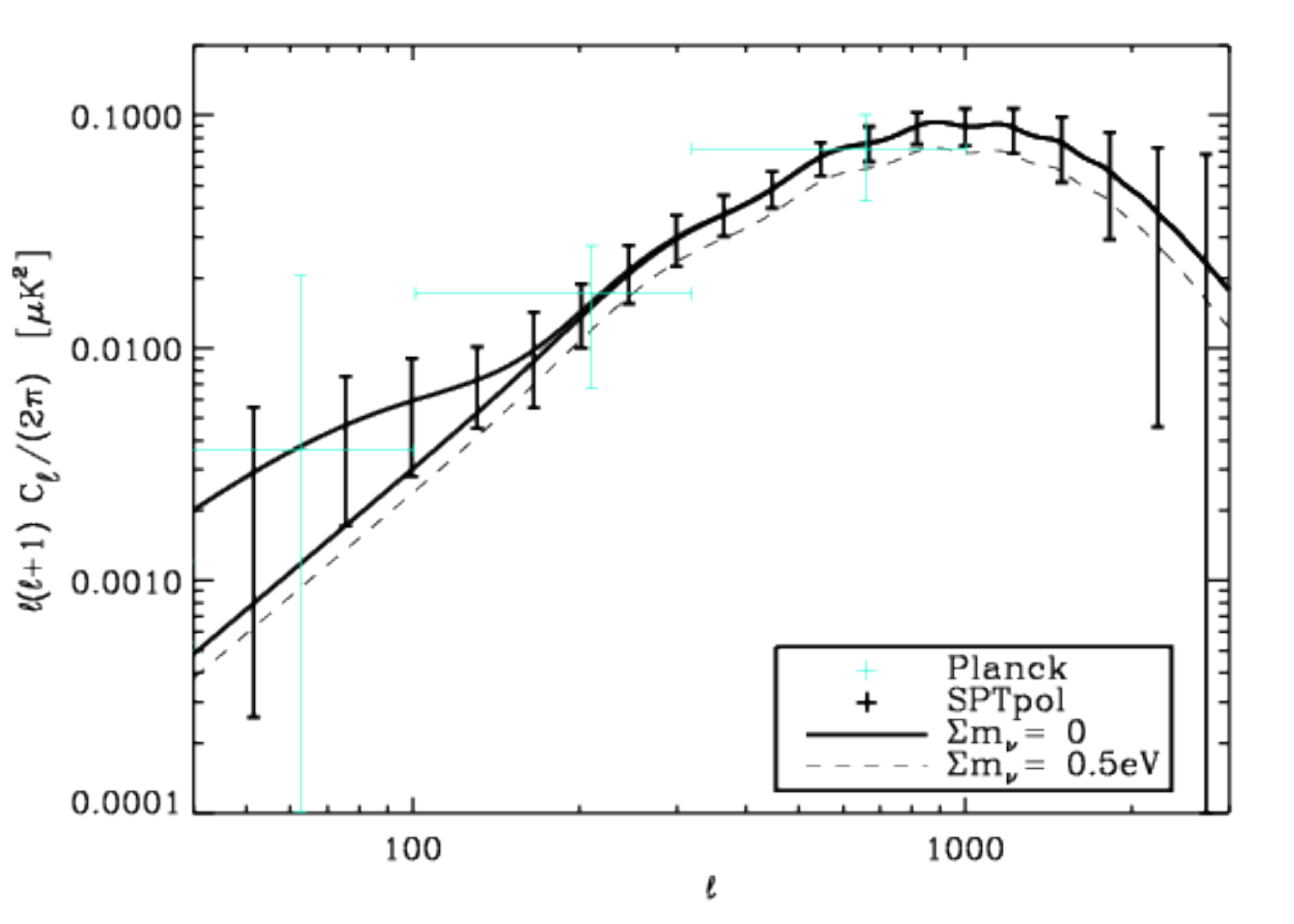}
\end{minipage}
\caption{
Projected $EE$ (left) and $BB$ (right) constraints from three years of observing with the SPTpol camera (black points and error bars).  
Constraints are from simulated observations including realistic treatment of foregrounds, atmosphere, instrument
$1/f$ noise, and $E$-$B$ separation while assuming detector performance listed in Table~\ref{tab:noise} with 80\% pixel yield.  
Overplotted are projected constraints  from \planck\ \cite{planck06}.  
Solid model curves in the $BB$ plot are for $\summnu=0$, with $r=0$ and
$r=0.04$, while the dashed model curve is for $\summnu=0.5eV$ with $r=0$.  
\label{fig:pspec}}
\end{figure}

\subsubsection{Inflationary B-modes}
The only process expected to generate primordial B-mode polarization anisotropy is from 
gravity waves generated during the inflationary epoch, which are predicted to 
produce a signal at large angular scales ($\ell < 100$) with an amplitude related to the energy 
scale of inflation, and proportional to the tensor-to-scalar ratio, $r$, of the primordial perturbations\cite{hu02b}.
Detection of this signal is often considered both a confirmation of the theory of inflation, and 
tantalizing information about physics at energy scales far beyond those probed at terrestrial accelerators\cite{weiss06}.

While the detection of gravitational-wave B-modes will be extremely
challenging, the raw sensitivity of SPTpol and the low-foreground sky
continuously observable from the polar site will allow SPTpol 
to place a cosmologically compelling constraint on the
gravitational-wave B-mode spectrum that could rule out various large-$r$ inflationary models.
The right panel of 
Figure~\ref{fig:pspec} shows simulated B-mode spectra for the scenarios of $r=0$ and $r=0.04$.  
Using the simulated observations described below, we project that with three years of observations
and realistic estimates of atmospheric contamination, 
the $1\sigma$ detection limit in the joint fit to gravity-wave and lensing template B-mode spectra is $r = 0.028$. 
This represents a significant improvement from the best current constraints on the tensor-to-scalar 
ratio from both the B-mode polarization, $r < 0.7$ at 95\% confidence\cite{chiang10}, and the CMB
temperature anisotropy, $r < 0.21$ at 95\% confidence\cite{keisler11}.

\subsubsection{Gravitational Lensing and Neutrino Masses}
The other significant source of B-mode polarization is generated from the gravitational lensing of the CMB by large-scale structure, 
which converts some of the intrinsic E-mode polarization to B-modes.  
This contribution to the B-mode angular power spectrum is expected to peak around $\ell \sim 1000$ and dominate 
the primordial B-mode signal at $\ell > 200$ for any 
allowed value of the tensor-to-scalar ratio $r$, for example, see Figure~\ref{fig:pspec}.
The shape and amplitude of the lensed B-mode power spectrum
depends on the sum of neutrino masses, \summnu, because the energy density of massive neutrinos
decreases more slowly with expansion than is the case for massless
neutrinos.  The resulting higher density, and thus
higher universal expansion rate, leads to slower growth of structure ---
and, hence, suppression of lensed B-mode power ---
on scales smaller than the neutrino free-streaming length.  
SPTpol, with its high-resolution $1^\prime$ beam, is capable of 
measuring the small-scale B-mode power spectrum with high precision,
resulting in scientifically interesting constraints on \summnu.
 
In the right panel of Figure~\ref{fig:pspec}, we show that the SPTpol measurement of the 
B-mode power spectrum has the power to clearly distinguish between spectra with 
$\summnu = 0$~eV and 
$\summnu = 0.5$~eV. 
Including priors from \planck, see Table \ref{tab:cosmology}, we expect to constrain $\sigma(\summnu) = 0.096$~eV, a constraint 
that is $\sim$4$\times$ better than future beta decay experiments such as 
KATRIN\cite{wolf10}, which has a predicted sensitivity of $\sim$0.6~eV (90\% confidence limit)
for \summnu\cite{gonzalez-garcia10}.  Moreover, the SPTpol constraints compliment other 
cosmological probes of the neutrino mass (e.g., galaxy clusters), which addresses potential 
uncertainties due to model dependence, and are comparable to the largest neutrino mass splitting of $\sim$0.05~eV.  

\subsubsection{Simulated Polarization Observations}
We estimate SPTpol constraints on the E-mode and B-mode power spectra by
performing Monte-Carlo
simulations of three years of SPTpol observations that uniformly map a $625 \ \sqdeg$ field.  
The simulations assume a realistic scan strategy with 1 deg~$\mathrm{s}^{-1}$ azimuth scans, 
and a $50\%$ observing duty cycle during the Austral winter, which is
conservative compared to the $\sim 60 \%$ achieved by SPT-SZ.
The simulations include the effect of foreground contamination, which are mitigated through 
our choice of a relatively dust-free observing region, see Section \ref{sec:obs}, and suppressed further by making linear combinations 
of the two observing bands.  The simulations assume an instrumental $1/f$ knee at 0.05~Hz, and 
further suppression of atmospheric $1/f$ from pair differencing detectors in the same pixel.  
For this, we assume a frozen-screen atmospheric model and a factor of 200 
common-mode rejection from the differencing.  In SPT-SZ, we could achieve a factor of 100 reduction 
in atmospheric signal differencing neighboring detectors without any gain matching, and preliminary 
analyses of SPTpol indicate better performance\cite{Sayre_2012,Henning_2012}.
We use a version of the Smith et al. 2006 contaminant-free B-mode 
estimator\cite{Smith_2006} in these simulations
and find the residual E $\rightarrow$ B leakage 
to be below levels relevant for SPTpol, even when applied 
to non-cross-linked 
data that have been filtered to reduce the effects of $1/f$ noise. 
We fit the B-mode spectra jointly to a gravitational-wave template
spectrum and a lensing template spectrum and estimate joint 
constraints on $r$ and \summnu.
We note that effects such as polarized sidelobe pickup,
beam imperfections, and calibration uncertainties are not included
in the simulations, but are expected to be sub-dominant to the statistical errors of this survey
due to mitigation techniques of design and strategy outlined in Section \ref{sec:obs}.  

\begin{table}[t]
\vskip 12 pt
\small
\begin{center}
\begin{tabular}{l | ccccccccccc}
\hline
\multicolumn{1}{c|}{Dataset} 
& \multicolumn{9}{c}{Cosmological parameter constraints} \\                  
  & $\sigma(\ombh)$ & 
  $\sigma(\omch)$ & 
$\sigma(A_s)$ & 
$\sigma(n_s)$ & 
$\sigma(h)$ & 
$\sigma(\tau)$ & 
$\sigma(\neff)$ & 
$\sigma(\summnu)$ & 
$\sigma(r)$ \\
  &
$\times 10^{4}$ & 
$\times 10^{3}$ & 
$\times 10^{11}$ & 
$\times 10^{3}$ & 
$\times 10^{2}$ & 
$\times 10^{3}$ & 
$\times 10^{1}$ & 
[meV] & 
$\times 10^{2}$
\\\hline
\planck{}   &  
{1.93} & 
{ 2.02 } & 
{ 5.36 } & 
{ 7.07 } & 
{ 1.88 } & 
{ 4.96 } & 
{ 1.39 } & 
{ 117 } & 
{ 5.72 } \cr
~~~~+\,\normalsize SPTpol{} \small&  
{ 1.64 } & 
{ 1.71 } & 
{ 4.92 } & 
{ 6.19 } & 
{ 1.58 } & 
{ 4.95 } & 
{ 1.17 } & 
{ 96 } & 
{ 2.75 } \cr

\end{tabular}
\vspace{0.1in}
\caption{Expected $1\,\sigma$ constraints on cosmological parameters using SPTpol 
power spectrum and lensing reconstruction data, assuming a 9-parameter \lcdm+\neff+\summnu+tensor model. 
In addition, when applied to a model including the primordial helium abundance (\lcdm+\neff+$Y_p$+\summnu\,  cosmology) SPTpol will improve the constraint on $Y_{\rm{He}}$ by a factor of $\sim 2$ over \planck\ alone.
}
\label{tab:cosmology}
\end{center}
\end{table}

\subsection{E-mode and Fine Scale CMB Anisotropies}
The primary E-mode polarization signal is generated from the same acoustic oscillations as those that source the temperature fluctuations in the CMB, but is sourced from the velocity field rather than the temperature of the plasma.  The E-mode power spectrum ($EE$) is consequently out of phase with the density fluctuations that power the temperature power spectrum ($TT$).  This means the $EE$ power spectrum carries much the same information as the $TT$ spectrum.  Although the E-mode signal is generally at least a factor of 6 weaker that the temperature signal, it has the distinct advantage of suffering from far weaker foreground contaminates (e.g. dusty galaxies, galaxy clusters, galactic sources) that become important at high-$\ell$ and are difficult to separate in the $TT$ spectrum\cite{Shirokoff_2011}.  For example, at 150~GHz, the dominant high-$\ell$ foreground contaminants are dusty galaxies, which are expected to be polarized at only the 1--2\% level\cite{seiffert07}.  Therefore, at the limit of systematic uncertainty from foreground contamination dominating statistical uncertainty, the $EE$ spectrum can provide improved constraining power on cosmological parameters sensitive to the primordial high-$\ell$ spectrum.

The primordial fluctuations at high-$\ell$ are sensitive to the primordial helium abundance, $Y_{\rm{He}}$, through its effect on the epoch of recombination.  The density of helium affects the electron density during recombination such that, for a given baryon density, a larger helium abundance will result in a lower electron density during recombination, which results in diffusion damping on larger scales (suppression of high-$\ell$ acoustic peaks).  The effects of helium density on electron density during the epoch of recombination also results in small shifts in the location of the acoustic peaks at high-$\ell$\cite{ichikawa08}.  The relatively clean measurement of the primordial high-$\ell$ spectrum provided by E-modes will allow significant improvements on the measurement  of $Y_{\rm{He}}$, which leads to an independent measurement of the baryon density ($\Omega_{\rm{b}}h^2$) through Big Bang Nucleosynthesis (BBN) and helps break the degeneracy between $\Omega_{\rm{b}}h^2$ and $n_{\rm{s}}$, the scalar spectral index.

SPTpol will also expand on much of the fine-scale CMB anisotropy work done by SPT-SZ.  
SPTpol will probe secondary CMB anisotropy due to the background
of lower mass SZ clusters below the detection threshold\cite{lueker10}  
and the clustering power of the cosmic infrared background (CIB) in the millimeter\cite{Hall_2010, shirokoff11}.
Measurements on these scales will also provide constraints on the evolution 
of the ionized fraction during the epoch of reionization through the kinetic SZ effect \cite{reichardt11,zahn11b}.
Comparing the gravitational lensing of the CMB to galaxy surveys or other external mass tracers provides another 
source of constraint on dark energy and structure formation 
\cite{Bleem_2012b,vanengelen12}.
Fine scale anisotropies are also sensitive to the expansion rate during recombination and thus to the number of relativistic particle species present at that epoch.  This has allowed SPT to place the tightest yet constraints on the number of light particle species beyond the standard three neutrinos\cite{benson11} (e.g., sterile neutrinos).
 
\subsection{Clusters of Galaxies}

Clusters of galaxies are the largest gravitationally bound objects in the Universe.  Their large masses make them a
unique cosmological probe sensitive to gravity and the growth of structure on the largest physical scales.  
As demonstrated by SPT-SZ, a high-resolution SZ cluster survey can uniquely find the most massive clusters in the Universe 
nearly independently of redshift\cite{reichardt12}.  Relative to SPT-SZ, SPTpol will have a factor of $\sim$1.3 times
lower mass threshold and find a comparable number of clusters, with the expected number of clusters per \sqdeg\ being 
a factor of $\sim$3 times larger.  The lower mass threshold will also extend the redshift reach of the cluster survey, and 
test the effect of dark energy and neutrino mass on the growth of structure at even higher redshifts.

\subsection{Dusty Star Forming Galaxies}
The SPTpol survey will also discover a significant number of faint extragalactic discrete sources at mm-wavelength.
The large areas and high sensitivities of the SPT-SZ and SPTpol surveys are ideal for 
detecting the brightest mm-wavelength objects in the Universe.  These bright mm-wavelength sources tend to be 
strongly lensed dusty star forming galaxies (DSFGs)\cite{vieira10}, massive, dusty galaxies that make up a significant 
fraction of the high-redshift ($z > 1$) component of the cosmic infrared background (CIB), and are crucial to our 
understanding of galaxy formation.  
This capability makes the SPTpol survey an important compliment to the high spatial and spectral resolution of ALMA, by 
identifying the most extreme mm-wavelength bright sources in the Universe.  
With ALMA, the high-redshift lensed structures will be useful measures of star formation near the end of the epoch of reionization.
They will also probe the dark matter distribution in the elliptical galaxy lenses, to characterize mass distributions, substructure abundance, and determine the mass-to-light ratio in elliptical galaxies.

\section{Telescope and Optics} 
\label{sec:tel}

\begin{figure}[t]
\centering
\begin{minipage}[c]{0.49\textwidth}
\includegraphics[trim=0pt 0pt 0pt 0pt,width=0.99\textwidth]{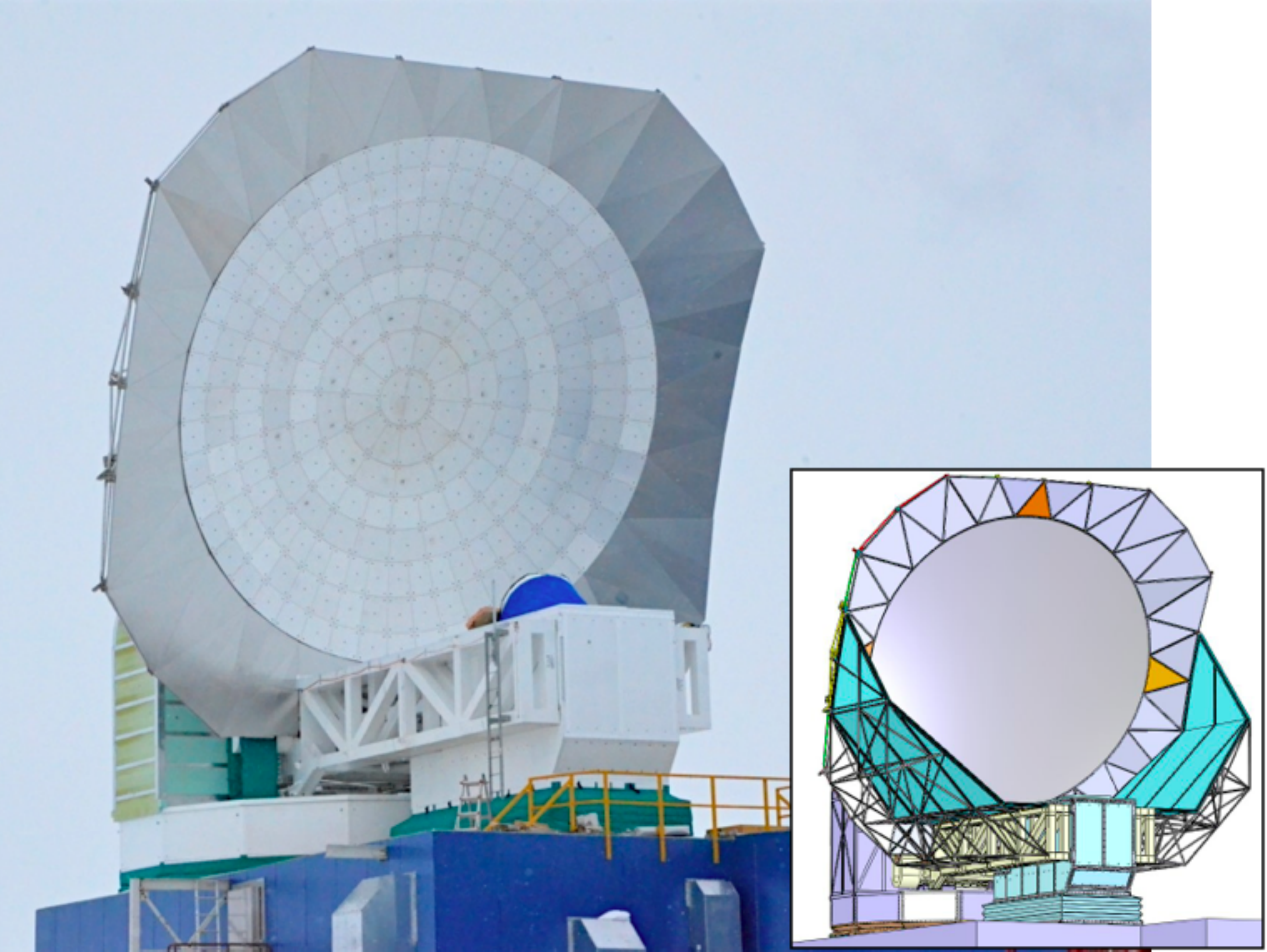}
\end{minipage}
\begin{minipage}[c]{0.49\textwidth}
\includegraphics[trim=0pt 0pt 0pt 0pt,width=0.99\textwidth]{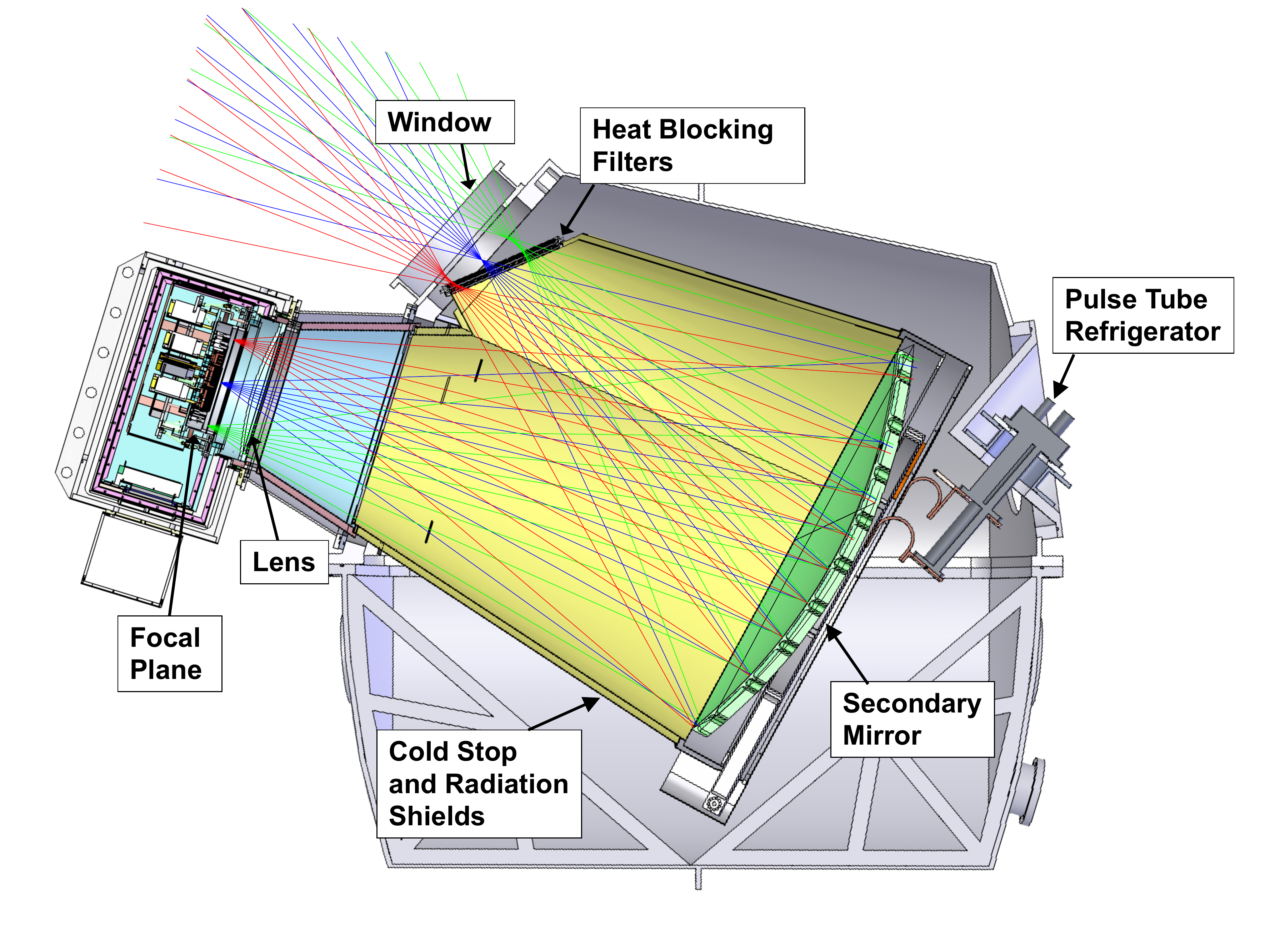}
\end{minipage}
\vspace{.1in}
\caption{\label{fig:tel} {\it Left}: Photograph of the SPT with the new RF shield 'guard ring' 
extended past the 10 meter primary, installed during the 2011/12 Austral summer.  
The inset shows the mechanical design of the second component of new shielding to be installed in November 2012.
{\it Right}: Cross section of the optics and receiver cryostats with representative beams from the center and edges of the focal plane.
}
\vspace{.1in}
\end{figure}

The SPT is a 10~meter telescope optimized for sensitive, high-angular resolution  measurements of the anisotropy of the  
CMB and mm-wavelength sky \cite{Padin_2008, Carlstrom_2011}.
The telescope is located at the NSF Amundsen-Scott South Pole station,
the best location on Earth for millimeter-wave observations, with 30 times less atmospheric fluctuation 
power than found at the ALMA site in the Atacama desert \cite{Bussmann_2005, Radford_2011}.
The telescope is an off-axis, classical Gregorian design that gives a 
wide diffraction-limited field of view, low scattering and high efficiency
with no blockage of the primary aperture. 
The current telescope optics produce a $1^{\prime}$
FWHM beamwidth at 150~GHz with a conservative illumination of the inner 
8 meters of the telescope, and a $\sim$1~deg$^2$ diffraction-limited field of view\cite{Padin_2008}.
The SPT is designed to modulate the beams on the sky by slewing the entire telescope
at up to 4~deg~s$^{-1}$ and eliminating the need for a chopping mirror. The telescope operates largely remotely, with a high observing efficiency. 
In the Austral summer of 2011/2012, the ground shielding of the telescope was improved by 
adding a $\sim$1.5~m wide ``guard-ring" around the 10~m diameter primary mirror , see Figure \ref{fig:tel}.  
In November 2012, additional shielding will be installed along the boom of the telescope for further improvements. 

The SPT has a relatively simple optical design 
with just two mirrors (a primary and secondary) and one lens, with detectors located at the Gregory focus
, see Figure \ref{fig:tel}.  From the sky, light enters through a zotefoam vacuum window, 
which has $>$99\% transmission averaged across both the 90 and 150 GHz bands.  Directly behind the window 
are a series of infra-red shaders and blockers at $\sim$100 and $\sim$10~K, which are tilted to direct reflections 
away from the focal plane.  After the filters, light is reflected off a 1~m secondary mirror, which also effectively 
acts as an optical stop for the system.  For this reason, the secondary is cooled to 10~K with the optical path 
between the infra-red blocking filters, to the secondary, to the detectors, surrounded by HR-10 microwave absorber 
also cooled to 10~K.  The majority of this optics is housed inside a separate cryostat from the receiver with 
its own pulse tube refrigerator.  Next, another infrared blocking filter sits on top of a lens which re-images
the light at the Gregory focus to make a tele-centric focal 
plane.  The lens is made of high density polyethylene (HDPE) with an expanded 
teflon anti-reflection coating designed to keep reflections at $< 1$\% averaged across the 90 and 150 GHz bands.  
The lens is housed in the SPTpol receiver cryostat, and is cooled to $\sim$6~K.  Next, 
a series of low-pass metal-mesh filters are located near the focal plane, to reduce optical loading on the 
detector cold stage (280~mK), and define the high-end of the detector band-pass.

\begin{figure}[t]
\centering
\begin{minipage}[c]{0.28\textwidth}
\includegraphics[trim=0pt 0pt 0pt 0pt,width=0.99\textwidth]{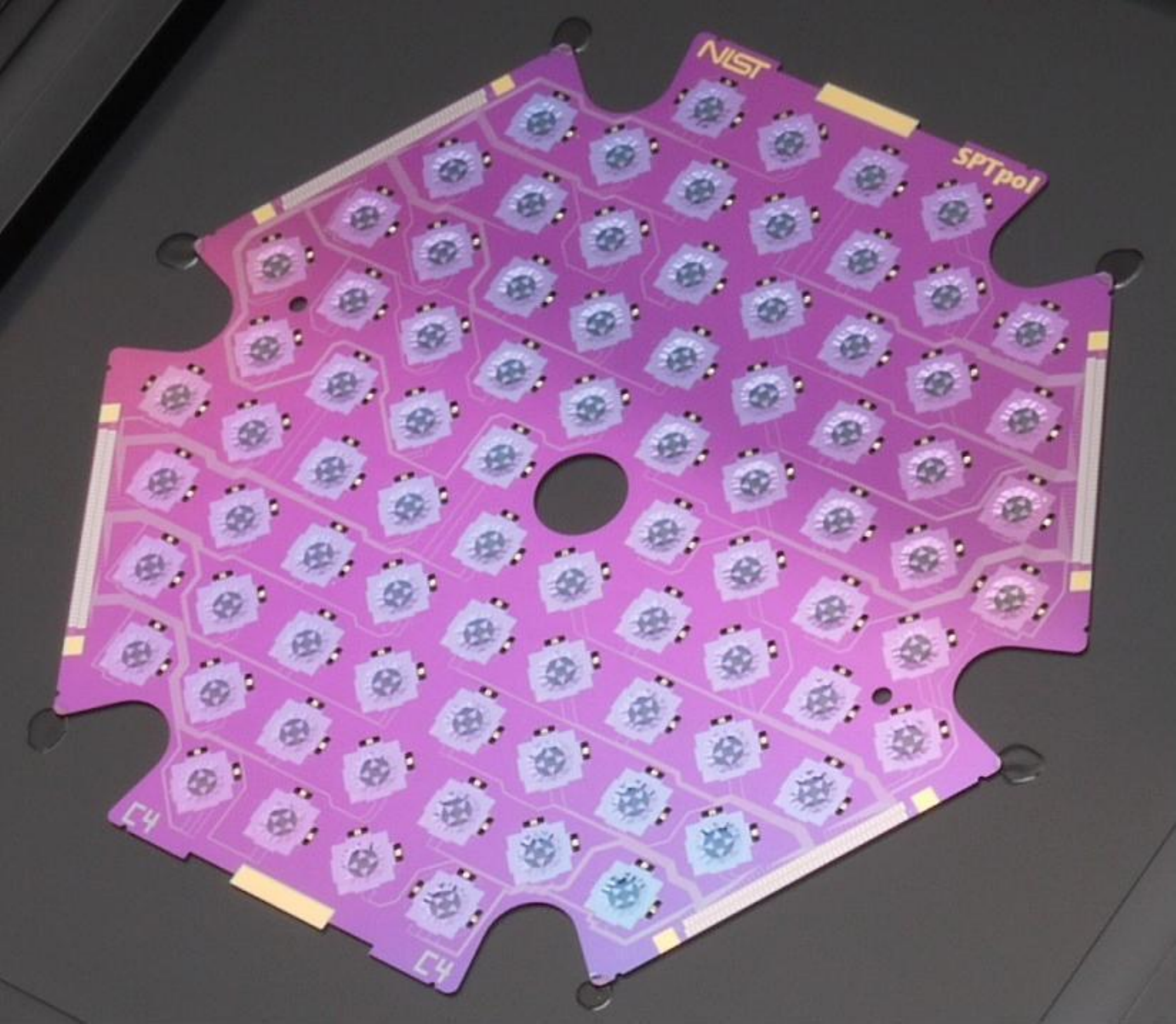}
\end{minipage}
\begin{minipage}[c]{0.285\textwidth}
\includegraphics[trim=0pt 0pt 0pt 0pt,width=0.99\textwidth]{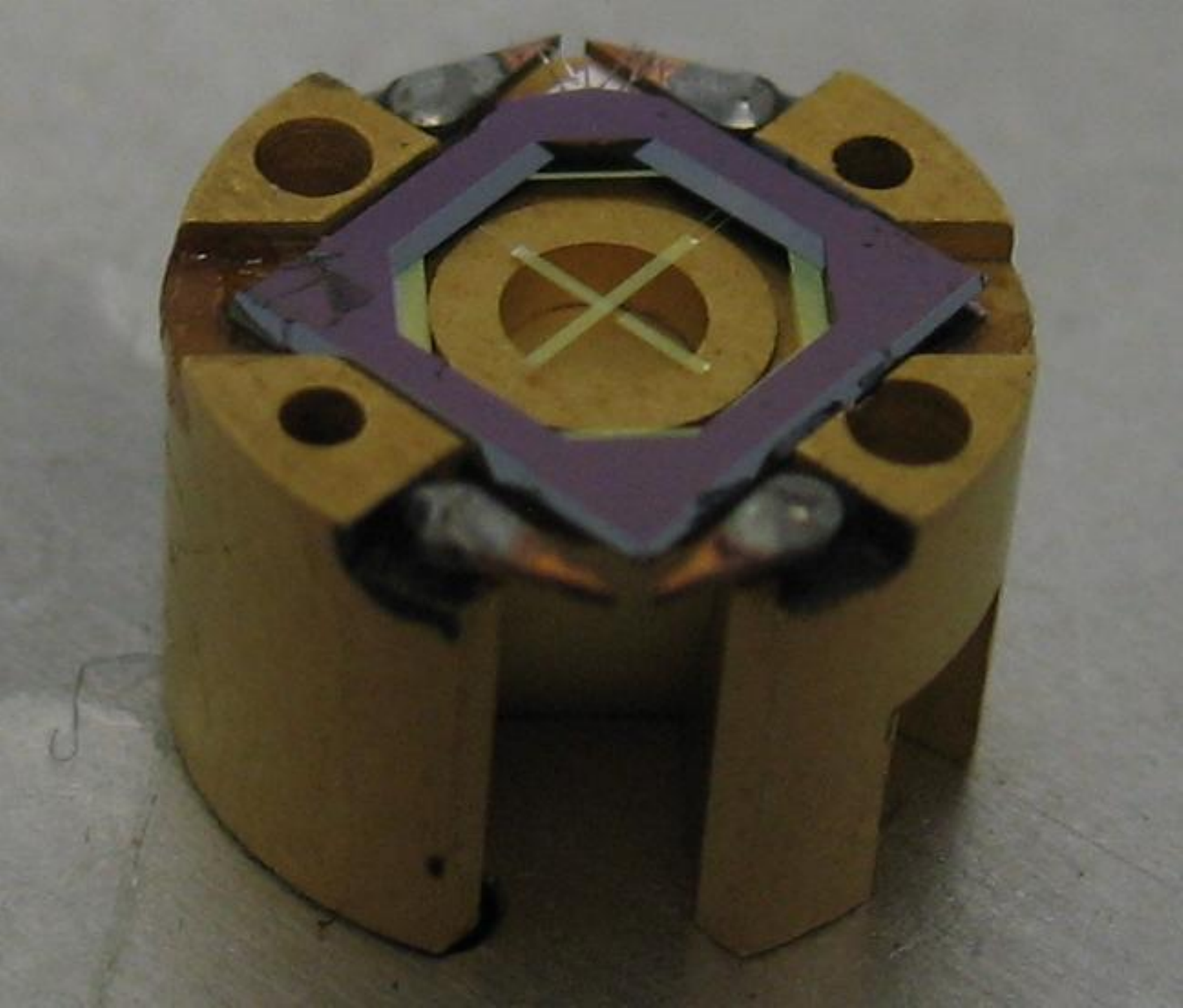}
\end{minipage}
\begin{minipage}[c]{0.30\textwidth}
\includegraphics[trim=0pt 0pt 0pt 0pt,width=0.99\textwidth]{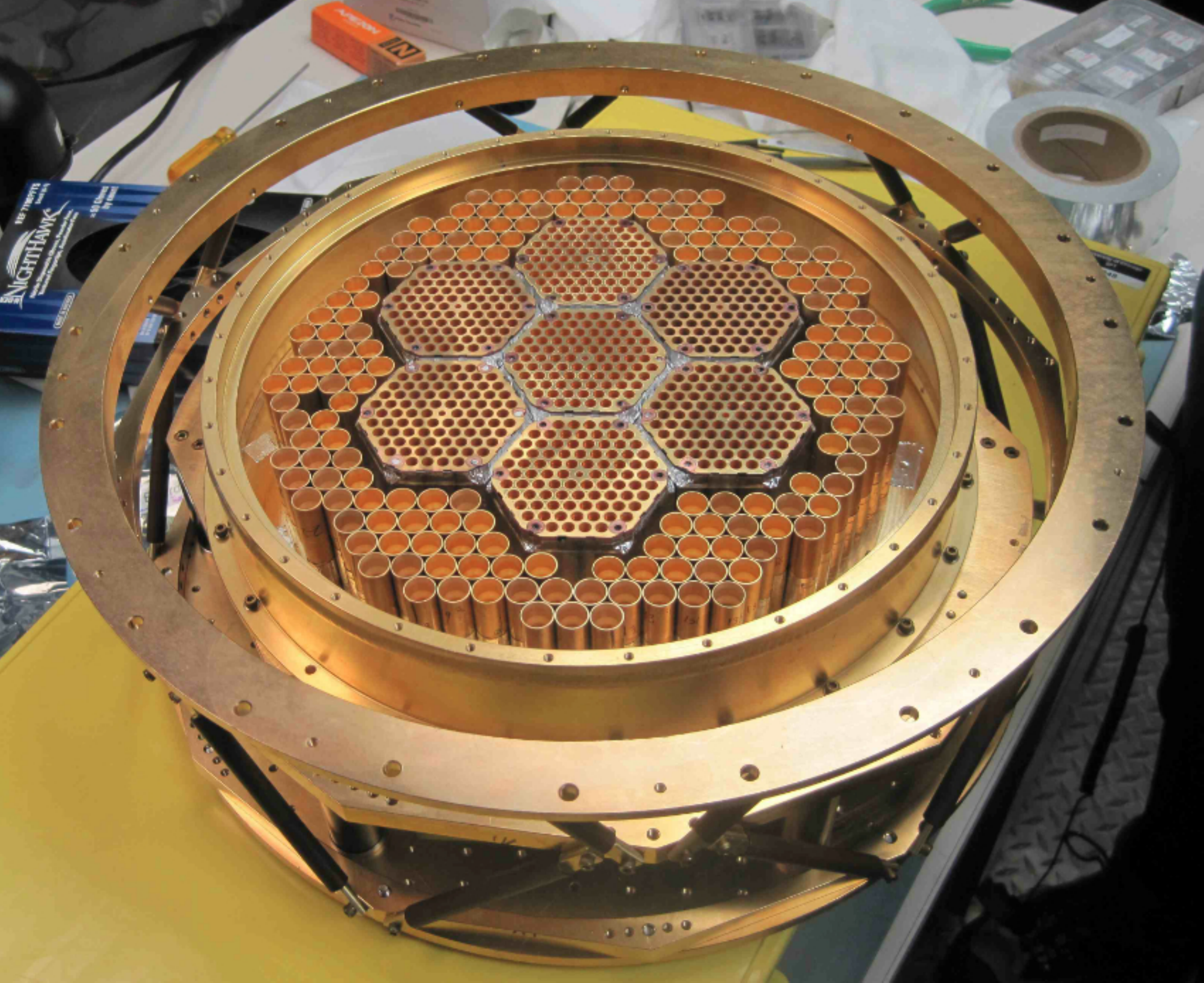}
\end{minipage}
\vspace{.1in}
\caption{\label{fig:det} {\it Left}: Monolithic 150~GHz array, consisting of 84 polarimeters (comprising 168 optical and 5 dark TES bolometers). 
Pixel orientations vary between 0 and 45$^\circ$ from nominal in order to improve polarization crosslinking when scanning.
{\it Center}: Single 90~GHz pixel in module with feedhorn and waveguide removed.  
{\it Right}: Photograph of the SPTpol focal plane populated with 150~GHz array modules (center, with smaller aperture feedhorns) 
and individual 90~GHz pixel modules (outer rings of feedhorns).}
\vspace{0.1in}
\label{fig:pix}
\end{figure}

\section{SPTpol Receiver} 
\label{sec:rec}

\subsection{Focal Plane} 

The SPTpol focal plane is filled with 768 feedhorn-coupled, dual-polarization pixels, each consisting of two transition-edge sensor (TES) bolometers for simultaneous detection of incident power in each of two orthogonal polarizations (for a total of 1536 optically coupled TES bolometers).  The focal plane area is split nearly evenly between 90 and 150~GHz sensitive pixels, with the outer ring of the focal plane consisting of 180 pixels at 90 GHz and the inner diameter consisting of 588 pixels at 150 GHz, see Figure \ref{fig:pix}.
The 90 and 150~GHz pixels were developed independently at Argonne National Laboratory (ANL) and the 
National Institute of Standards and Technology (NIST) in Boulder, CO, respectively, and are briefly reviewed below with details found elsewhere in 
these proceedings\cite{Sayre_2012,Henning_2012}.

\subsubsection{90~GHz Polarimeters}

Each 90~GHz pixel is built as a single module that comprises two individual, but identical, single-polarization detectors mounted face-to-face 
and rotated 90~degrees with respect to each other.  A contoured feedhorn with a single-moded circular waveguide couples light 
to the detectors via a resistive PdAu absorbing bar, which is connected to a
SiN thermal mass with a lithographed Mo/Au bilayer TES.  
In this configuration, the 90~GHz modules have been measured in the lab to have low cross-polarization response ($< 1.6$\%), 
high optical efficiency for each polarization ($\sim 87$\%), and excellent noise properties consistent with expectations\cite{Sayre_2012}.
The 90~GHz modules are built individually and can be installed to the focal plane as independent units, thus allowing replacement 
of broken pixels or those with poor performance.  A 90~GHz pixel with the feedhorn and waveguide removed can be seen in Figure~\ref{fig:pix}.

\subsubsection{150~GHz Polarimeters}

The SPTpol 150~GHz polarimeter design is based on the generic mm-wave polarimeter development 
by the TRUCE collaboration \cite{Yoon_2009,Austermann_2009,Bleem_2009} that is being used in multiple upcoming and future CMB experiments
including ABS\cite{essinger10} and ACTpol \cite{Niemack_2010}.  
SPTpol detectors represent a branch of this development where TES designs were modified to match the 
expected loading and readout requirements of the SPTpol experiment.  The layout was expanded into a monolithic 
array of 84 polarimeters, and on-chip passband filters
were removed in favor of low-pass free space filters and a high-pass waveguide in order to 
improve optical efficiency and block out-of-band leakage \cite{Henning_2010}.

Optical power is coupled to the polarimeters through
corrugated feedhorns, which help provide low cross-polarization, symmetric beams,
high transmission efficiency, and wide bandwidths. 
Corrugated, gold-plated, silicon platelet feedhorn arrays were developed at
NIST \cite{Hubmayr_2012a} and made to match the 84-pixel polarimeter arrays. 
Single-moded waveguide feeds incoming radiation to a
planar orthomode transducer (OMT) that separates
the radiation into orthogonal polarizations.  Each polarization is then coupled
through micro-strip to its own TES bolometer island,
where the conductor transitions from superconducting
into a lossy gold meander that dissipates the radiation
into heat measured by the TES.  The 150~GHz passband is 
defined through a series of low-pass free-space filters located throughout the optical chain
and the high-pass cutoff frequency of the waveguide section between the feedhorns and detectors.
Lab tests of the feedhorn and detector arrays exhibit excellent cross-polarization (below $-25$ dB) and
optical efficiency ($\sim 90$\%) properties, with measured noise levels consistent with predicted design
levels \cite{Henning_2012}. In addition, the monolithic design results in consistent alignment of pointing and polarization
between the two polarization channels of each pixel.


\begin{table}[t]
\small
\begin{center}
\begin{tabular}[c]{cccccccc}

\hline
Band  & Number of Pixels & Resolution & Horn Diam. & Detector \& Horn &  End-to-end     & Target Survey depth \\
(GHz) & (polarimeters) &  (arcmin)  & (f$\lambda$)   &  optical eff.  &  optical eff.  &  \ukarcmin\ \\ 
\hline
90    & 180 (360 TES)  &  1.6       &  2.3  &  $\sim$0.87                & $\sim 0.46$            &   $9$ \\
150   & 588 (1176 TES) &  1.0       &  1.6  &  $\sim$0.90         & $\sim 0.45$            &   $5$ \\
\hline

\end{tabular}
\vspace{0.1in}
\caption{Base parameters for the SPTpol polarimeter and projected sensitivity of the full 625~\sqdeg\ SPTpol survey.  Each polarimeter comprises two detectors, for a total of 1536 optically coupled detectors in the focal plane. Detector--horn optical efficiency have been confirmed through lab and in situ measurements \cite{Henning_2012,Sayre_2012}. System optical efficiency is calculated for a single polarization from a combination of measured and expected efficiencies of all optical components and has been confirmed through observation \cite{George_2012}. 
Depth numbers are projected 
noise rms (in T; Q and U will be $\sqrt{2}$ higher) in $1^\prime$ pixels for the full 4 years of 
observation on a 625~\sqdeg\, field and are calculated using achieved mapping speeds and efficiencies of SPT-SZ scaled to the SPTpol 
sensitivities and yield (Table~\ref{tab:noise}). } 
\label{tab:basics}
\end{center}\vskip -0.1 in
\end{table}

\subsection{Readout} 

All detectors are biased and read out using a second-generation digital
frequency-domain multiplexer system with cryogenic SQUID amplifiers based on
the initial SPT-SZ readout \cite{Smecher_2012, Dobbs_2012}. 
This low-noise system introduces a small current-noise term that is sub-dominant to 
other noise sources under normal operating parameters \cite{George_2012}. 
For SPTpol we multiplex 12 resonant channels on each SQUID \cite{Henning_2012}, 
with frequency separation between channels of $\gtrsim 60$~kHz to minimize crosstalk.  
A total of 144 SQUIDS are used to readout the 1536 optical TES channels
in addition to various dark TES bolometers and other calibration channels.
Further details on SPTpol readout electronics, control software, data acquisition, 
data handling and archiving can be found elsewhere in these proceedings \cite{Story_2012}.

\section{Observing Strategy} 
\label{sec:obs}

\begin{figure}[t]
  \center{\includegraphics[height=7cm]
    {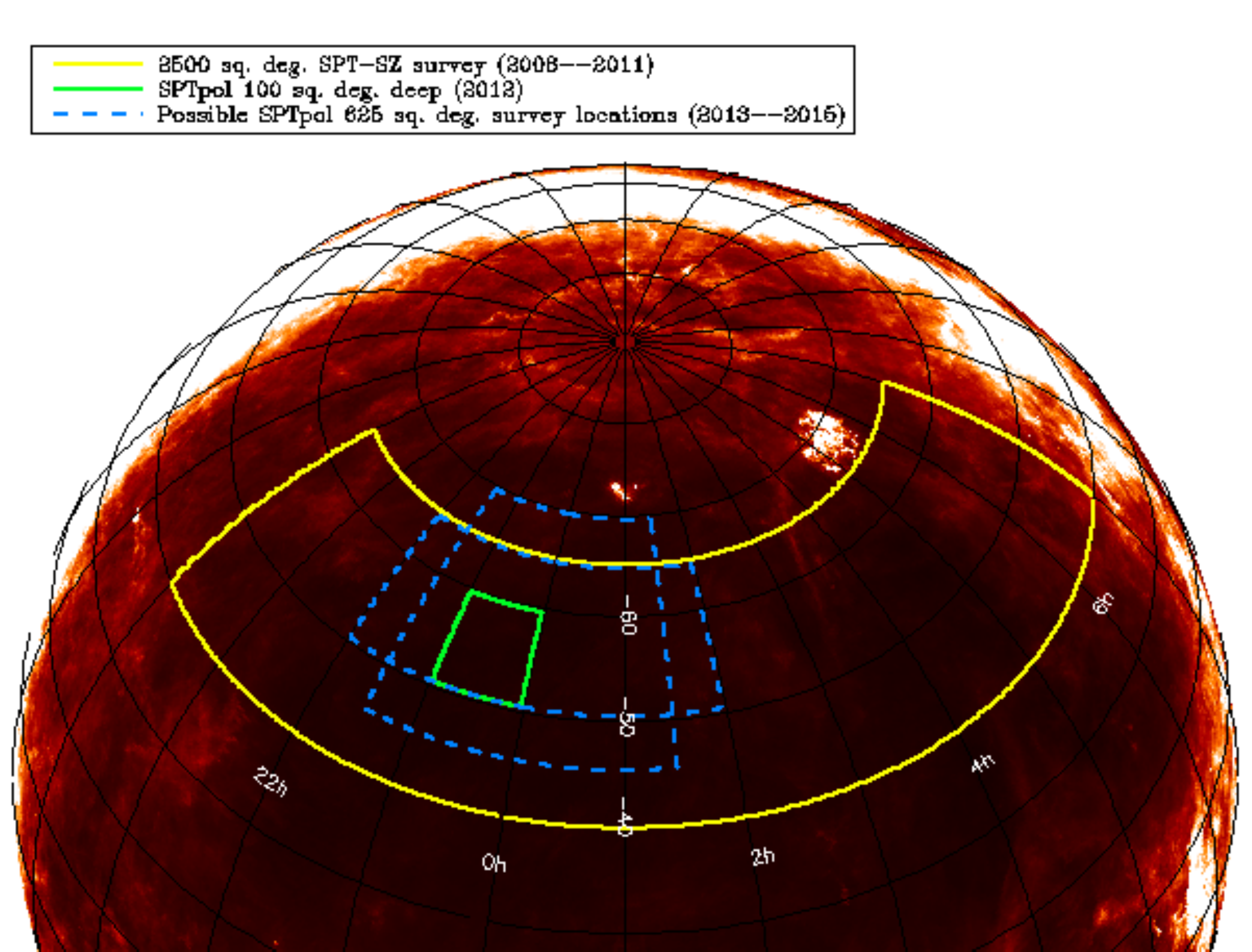}}
  \vspace{.1in}
  \caption{\label{fig:fields} SPTpol survey field locations relative to SPT-SZ plotted over-plotted on an IRAS 100 \micron\, dust map\cite{Schlegel_1998}. 
  The sky is rotated such that the South Celestial Pole is at the top of the globe, and R:A: = 1h faces the viewer }
  \vspace{.1in}
\end{figure}

The SPT was designed to conduct large-scale surveys at millimeter and sub-millimeter wavelengths, 
taking advantage of the exceptionally stable and transparent atmosphere above the South Pole. 
The unique atmosphere at the South Pole site enables relentless observing in remarkably 
stable conditions. Furthermore, the geographical location allows the survey observations 
to be conducted 24 hours-a-day, 7-days-per-week, year-round on a single region of sky through 
the same airmass: sources do not rise or set. This allows extremely deep, large-area 
surveys of the southern sky, which includes some of the lowest foreground and most studied CMB fields available (Figure~\ref{fig:fields}),
to be completed quickly and with highly uniform sensitivity. 
Accordingly, all other aspects of the SPTpol project---the telescope, the RF shielding, the receivers, 
the observing strategy and survey design---have been optimized to minimize systematics and allow
ultra-sensitive measurements of the CMB.

\subsection{Survey Parameters and Overlap With Other Data Sets}

Over a 4 year period that started in February, 2012, SPTpol will map approximately
625~\sqdeg to expected depths of $\sim 9$ and $\sim 5$ \ukarcmin\ at 90 and 150~GHz, respectively.  
The SPTpol survey is located in a region of low galactic foreground
and will overlap with the BICEP2 and KECK CMB fields.  
The exact location and dimensions of the full survey have not been finalized at this time; however, likely locations are depicted in Figure~\ref{fig:fields}. 

In advance of the cosmological constraints from the full SPTpol 
survey, we will leverage previous SPT-SZ data and a wealth of multi-wavelength
data to maximize early science from SPTpol in the first year of 
observation.  The entire first year will be spent observing the SPT-SZ 100~\sqdeg\
field centered at right ascension 23$^h$30$^m$, declination -55$^{\circ}$, located in 
one of the most foreground-free regions in the mm-wave sky
, see Figure~\ref{fig:fields}.  
This field is already the deepest large mm-wave field in the sky, 
having been observed with SPT-SZ to $\sqrt{2}$ below the normal SPT-SZ survey 
noise levels, and SPTpol will image it to a factor of $\sim$3 lower noise in just the first year.  
The multi-wavelength coverage of this field is unprecedented for a field this size.
It has already
been covered in the near-infrared and far-infrared/submm with dedicated programs 
using the \spitzer-IRAC and \herschel-SPIRE instruments; in the optical, the \des\
collaboration plans to observe this field to full survey depth in the first months 
after DECam commissioning; in the X-ray, a large fraction of the field will be covered  
with a dedicated \xmm\ program; and proposals are being considered
to cover the field in several other wavebands.

These multi-wavelength datasets, combined with deep mm-wave temperature
and polarization data, will 
enable several exciting scientific results related to 
cosmology and the growth of structure in the Universe.  The mass limit for 
SZ cluster detection in this field will be significantly lower than in the SPT-SZ
survey, allowing us to extend the SPT cluster catalog to lower mass and higher redshift.  This makes
the \spitzer/IRAC data in this field even more crucial, as measuring redshifts for 
these clusters will be very difficult in the optical but simple with deep IRAC data
\cite{brodwin10}.
The complimentary nature of SPTpol and \spitzer/IRAC cluster observations works in
the other direction as well: SPTpol will provide mass estimates for \spitzer-discovered
high-redshift clusters\cite{brodwin12}, as in Brodwin et al. 2012.  SPTpol data will similarly provide 
an initial mass calibration for \des-discovered clusters, and the combination of 
SPTpol, \des, and \spitzer-IRAC data will enable detailed studies of cluster and
galaxy formation at high redshift.  CMB lensing from the first 100~\sqdeg\ of 
SPTpol data will provide high signal-to-noise mass maps of this field, which 
can be correlated with tracers of large-scale structure from all the multi-wavelength
datasets, enabling measurements of galaxy bias at
low (\des), intermediate (\spitzer), and high (\herschel) redshift.  In addition, the combination of \herschel\ 
and SPTpol power spectra will improve constraints on the 
epoch of reionization.

\begin{figure}[t]
  \centering \includegraphics[trim=0pt 0in 0pt
  0.2in,width=1.0\textwidth]{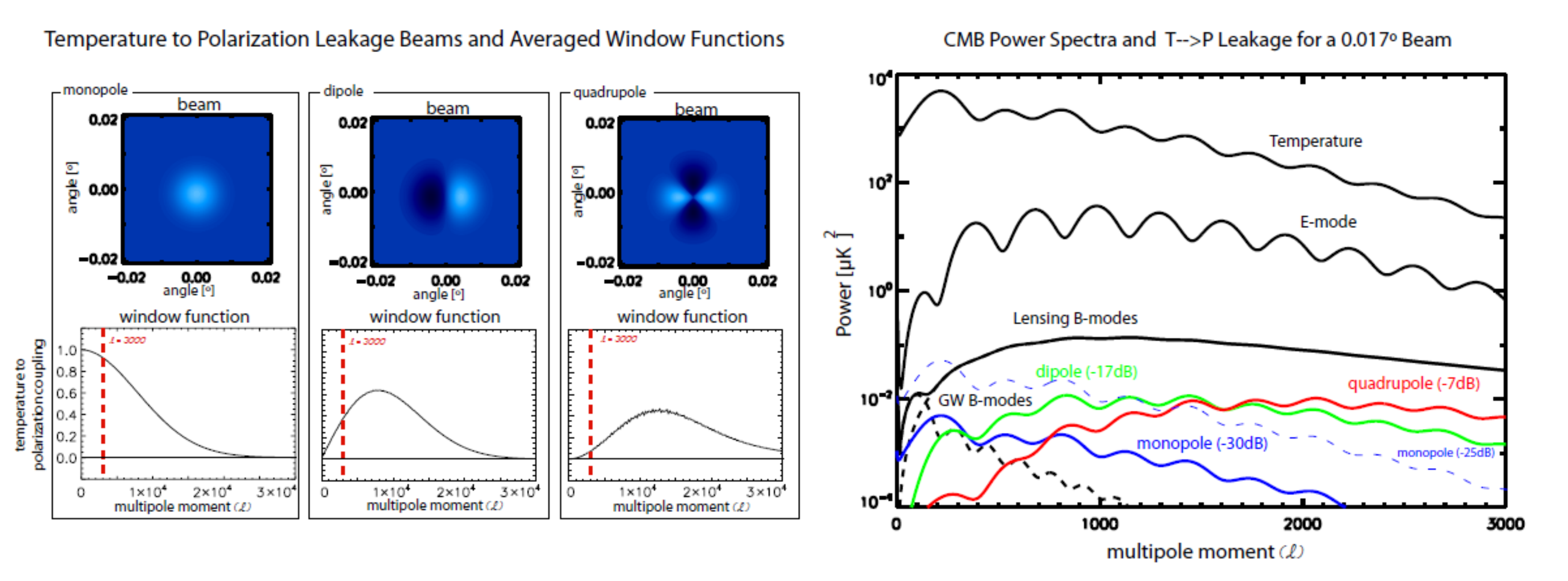}
  \caption{Effects of temperature to
  polarization ($T \rightarrow P$) leakage for the SPT beam size
  (1$^\prime$) and the requirements for leakage
  suppression. \emph{Left}: Beam shapes for monopole, dipole, and
  quadrupole $T \rightarrow P$ leakage and the corresponding
  azimuthally averaged window functions, which give the $T \rightarrow
  P$ leakage as a function of multipole moment $\ell$.  Note the
  dipole and quadrupole leakage are highly suppressed by the small SPT
  beam size where the B-modes peak below $\ell=3000$. Monopole
  leakage is accounted for with careful relative calibration, see
  Section~\ref{sec:sptpolcalibration}. \emph{Right}: Simulations of the
  CMB auto-correlation power spectra (Temperature, E-mode, and
  B-mode) and the leakage power spectra due to monopole (blue),
  dipole (green), and quadrupole (red) leakage when suppressed by
  $-30$~dB, $-17$~dB, and $-7$~dB respectively, which keep the $T
  \rightarrow P$ leakage at least $-10$~dB below the B-mode signal. 
  \label{fig:temptopolbeamsandwindws} }\vspace{0.00in}
\end{figure}

\subsection{Calibration and Mitigation of Polarization Systematics}
\label{sec:polsystematics}

CMB polarization experiments such as SPTpol build on the deep
experience gained from CMB temperature experiments, but the low signal level
and the requirement to maintain the fidelity of the
polarization pose new challenges.  
As with most current and planned bolometric CMB polarimeters, SPTpol will extract polarization signals by
simultaneous differencing of two detectors sensitive to orthogonal linear polarizations viewing
the same point on the sky and portion of the atmosphere.  In such a ``detector differencing" technique,
$T \rightarrow P$ leakage (``instrumental
polarization'') can be caused by mismatches in relative gain calibration, beam
shape, or bandpass in the two differenced detectors, and by relative pointing errors.
Detector polarization angle mis-calibration and pointing
reconstruction errors also cause E $\rightarrow$ B leakage
(``cross-polarization''). The SPTpol experimental design has
numerous features to control polarization systematics, including 
the large-aperture SPT 
telescope, a simple, well-shielded optical design, and careful calibration plans. 

\subsubsection{Large-Aperture Telescope} 
Although the gravitational wave B-mode signal peaks at degree angular
scales, the SPT large-aperture telescope, with
$1^\prime$ resolution, offers significant
advantages in reducing 
systematic
errors generated by beam asymmetries, such as beam shape mismatches in
differenced beams, which can cause $T \rightarrow P$ leakage
\cite{hu03}. As shown in the left panel of
Figure~\ref{fig:temptopolbeamsandwindws}, the 1$^\prime$ SPT beams are
sufficiently small that the differenced-beam asymmetries (characterized by beam
dipole and quadrupole moments) probe the damping tail of the CMB power
spectra where power is low.  
This attenuation of the $T
\rightarrow P$ coupling is a distinct advantage of SPT's small
beamsize, relative to other ongoing CMB polarization efforts.
Although precision beam characterization of SPTpol awaits the availability of
a planet to observe from Pole (e.g. Mars in September 2012), 
early results and the clean SPTpol corrugated feed design suggest  
the relatively mild requirements on beam shape mismatch
shown in Figure~\ref{fig:temptopolbeamsandwindws}
should be easily met.  

Another systematic that is mitigated by 
the high angular resolution of SPT is the
requirement on the beam dipole caused by pointing offsets between the two polarization 
beams from one pixel. Our goal is that the two
beams for a pixel be co-located to within about $2''$.  This should be 
achieved given the high precision of the relative pointing of the
detectors, and will be verified to a small fraction of the $1^\prime$ beam
when planet observations are available.

\subsubsection{Simple, Well-shielded Optical Design}
The SPT telescope is a very simple design consisting of just two
mirrors and one low-power lens.  This simple design minimizes spurious
polarization generation and distortion.  The design obeys the Dragone
condition giving zero polarization rotation (crosspol) at the center
of the field.  The two mirrors and lens give 0.03\% $T \rightarrow P$
leakage, which can be accounted for with proper calibration
(Section~\ref{sec:sptpolcalibration}), and optics thermal stability
(which is aided enormously by the lack of diurnal variation in the South Pole thermal 
environment).  The telescope is also designed to give high rejection of any
emission outside the main beam, achieved with
a combination of an off-axis telescope, cold stop at the secondary, 
and integral co-moving shield.

\subsubsection{SPTpol Calibration}
\label{sec:sptpolcalibration}
Precision measurements of faint polarized sources require special attention
to calibration and beam characterization.  
Absolute gain calibration for SPTpol will be obtained by comparing SPT-SZ CMB
temperature anisotropy maps to those produced by \wmap\ and \planck.  For sky coverage larger
than $500 \ \sqdeg$, it should be possible to achieve a 1--2$\%$
cross-calibration with \wmap\ and likely better with higher-sensitivity 
and higher-resolution maps from \planck.  As is being done for SPT-SZ, gain stability will 
be monitored by
periodically activating a chopped IR source viewed through a small
hole in the cold secondary,
with regularly scheduled elevation ``nods'' of the telescope that monitor
the calibration across the array by observing the
zenith-angle dependent atmospheric loading, and with several observations of Galactic sources each day.
The response patterns of 
each detector will be mapped using planets as
unpolarized (when unresolved) sources to calibrate the effects of $T \rightarrow P$
leakage in differenced detector pairs.

Absolute orientation angle of the detectors' polarization response 
must be calibrated to better than $0.5^\circ$ to
detect an $r = 0.01$ B-mode signal in the presence of the E-modes.
The Moon is perhaps the only well-understood, bright, stable and available polarized 
source in the SPTpol bands with polariztion known to this level, but would 
saturate the detectors to the point where they have little to no response.  
Although \planck\ and other instruments may continue to improve the catalog of
polarization calibration sources\footnote{For example, the polarization of Tau A is constrained to better than $1^\circ$ by combining \wmap\ with other measurements \cite{weiland10}; however, at a declination of $+22$, Tau A is not observable from the South Pole},
SPTpol will self-calibrate using a custom-built
polarized source located 3 km from the SPT in the far-field of the
telescope. 
The source is a chopped thermal black body polarized by wire grids located in the center of a large
reflector that acts to reduce the atmospheric loading on the detectors from the horizon. This system will provide an accurate, high signal-to-noise measurement of the polarization angle of every operational pixel.

\begin{figure}[t]
  \center{\includegraphics[height=5cm,natwidth=210,natheight=242]
    {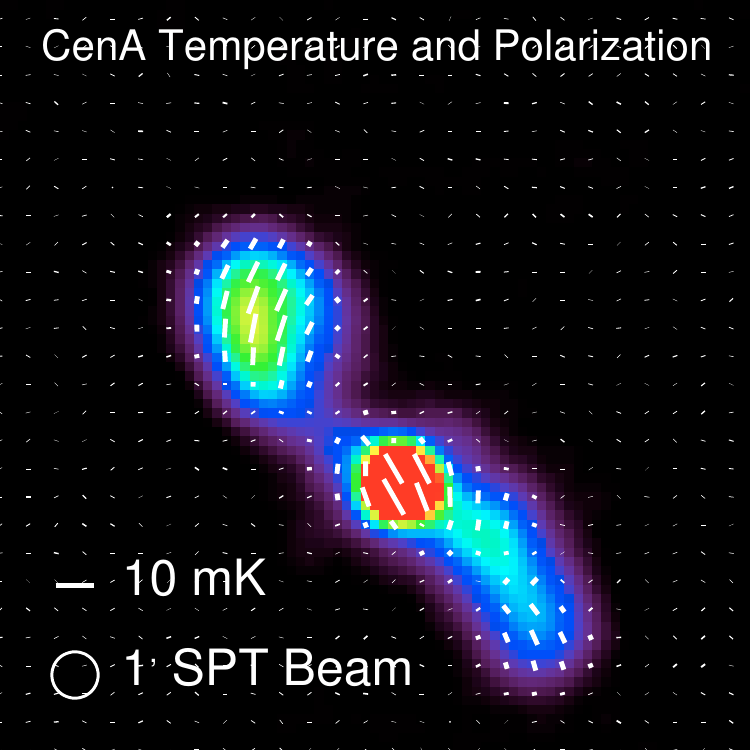}}
  \vspace{.1in}
  \caption{\label{fig:map} A preliminary image of Centaurus A at $150\,$GHz showing temperature (false color) and polarization (vectors), obtained during a 40 minute observation of a $\sim 1$ \sqdeg\ field during SPTpol commissioning in Feb. 2012.
    }
  \vspace{.1in}
\end{figure}

\section{Projections \& Performance} 
\label{sec:pro}

In Table~\ref{tab:noise} we give the measured band-pass, and the projected noise and sensitivity of an 
individual SPTpol TES bolometer during the 2012 observing season (Year 1), predicted using the measured detector 
parameters, band-passes, and under typical loading conditions.
Details of these measurements can be found elsewhere in these proceedings\cite{Henning_2012,Sayre_2012,George_2012}. 
Non-idealities in the Year 1 SPTpol band-passes\cite{George_2012} have lead to a noise performance that is slightly less than optimal.  
In Austral summer 2012/2013, the SPTpol filtering will be modified to optimize the high band-edge of both the 90 and 150~GHz 
bands, and the low band-edge waveguide cutoff of the 150~GHz pixels.  These modifications are expected to result in a 
$\sim$10\% improvement in the noise performance, and a $\sim$20\% improvement in mapping speed, in each band for 
observations in 2013 (Year 2) and onward, see Table~\ref{tab:noise}. 

The fraction of operational detectors and pixels (yield) during Year 1 was typically $\sim 80$\%.  This is expected to be improved 
in the following years from a combination of replacing cold and warm readout components, and fixing some bad cold wiring. 
At 80\% pixel yield, the Year 2+ combined focal plane NET is projected to be 26 and 15~\uk~${\sqrt {\rm s}}$ for 90 and 150~GHz, respectively.  
These sensitivities are used for the projected science results from the full 625~\sqdeg\ SPTpol survey, as discussed in Section~\ref{sec:sci} and summarized in Figure~\ref{fig:pspec} and Table~\ref{tab:cosmology}. 
The projected map depths for SPTpol also agree with a simple scaling from SPT-SZ using its measured focal plane sensitivity and survey map depth, 
which already include all overheads, data cuts, and observing inefficiencies, and would similarly predict final SPTpol survey depths
of $\sim9$ and $\sim5$\ukarcmin\ for the 90 and 150~GHz bands, respectively. 

\begin{table}[t]
\small
\begin{center}
\begin{tabular}[c]{ccccccccc}
\hline
Band & Center Frequency & $\Delta \nu $ & Optical Power & NEP$_{\rm Total}$ & NET$_{\rm }$\\
     & (GHz)      & (GHz)     & (pW)  & (aW / ${\sqrt{ \rm Hz}}$) & (\uk~${\sqrt {\rm s}}$) \\
\hline
\bf{YEAR 1} \\ 
\hline
90   &  91.2      &  30.1   & 10.2   & 95.5 & 501 \\
150  & 146.0      & 43.4     & 9.2    &  76.4 & 517 \\
\hline
\bf{YEAR 2+} \\
\hline
90   &  94.1      &  37.8    & 13.9   & 104.9 & 442 \\
150  & 148.5      &  45.5     & 8.22  & 73.1 & 469 \\
\hline

\end{tabular}
\vspace{0.1in}
\caption{Frequency band and noise performance estimates for the SPTpol receiver both before (Year 1) and after (Years 2+) anticipated band-pass upgrades.  Year 1 band-pass numbers were measured at the South Pole.  Noise estimates are given for an individual TES bolometer, predicted using the measured 
detector parameters, band-passes, and under typical loading conditions.   The predicted noise levels agree to within 
the 10--20\% systematic uncertainties of 
preliminary on-sky noise measurements\cite{George_2012}.
}
\label{tab:noise}
\end{center}\vskip -0.2 in
\end{table}

\acknowledgments    

Work at the University of Colorado is supported by the NSF through grant AST-0705302.  Work at NIST is supported by the NIST Innovations in Measurement Science program.  The McGill authors acknowledge funding from the Natural Sciences and Engineering Research Council, Canadian Institute for Advanced Research, and Canada Research Chairs program. MD acknowledges support from an Alfred P. Sloan Research Fellowship.  Work at the University of Chicago is supported by grants from the NSF (awards ANT-0638937 and PHY-0114422), the Kavli Foundation, and the Gordon and Betty Moore Foundation. Work at Argonne National Lab is supported by UChicago Argonne, LLC, Operator of Argonne National Laboratory (``Argonne''). Argonne, a U.S. Department of Energy Office of Science Laboratory, is operated under Contract No. DE-AC02-06CH11357. We acknowledge support from the Argonne Center for Nanoscale Materials.


\bibliographystyle{spiebib}   
\bibliography{austermann2012.bib}   

\end{document}